\documentclass[prb,aps,showpacs,a4paper,floatfix]{revtex4}
\usepackage{amsmath,textcomp,graphicx}
\usepackage{times}
\begin{document}
\title{%
(Ga,In)P: A standard alloy in the classification of phonon mode behavior
}
\author{O. Pag\`{e}s,\footnote{%
Author to whom correspondence should be addressed.
Electronic Address: pages@univ-metz.fr}
A. Chafi, D. Fristot, and A.V. Postnikov}
\affiliation{Laboratoire de Physique des Milieux Denses, Universit\'{e} 
de Metz, 1 Bd. Arago, 57078 Metz, France}
\begin{abstract}
Contrary to a broadly accepted assumption we show that random 
(Ga,In)P is not an exception with respect to the crude classification 
of the phonon mode behavior of random mixed crystals in terms of 
1-bond$\rightarrow$1-mode systems or 2-bond$\rightarrow$1-mode systems, 
as established from the simple criterion derived by Elliott \textit{et al.} 
[R.J. Elliott \textit{et al.}, Rev. Mod. Phys. \textbf{46}, 465 (1974)]. 
Consistent understanding of the puzzling Raman/infrared behavior of (Ga,In)P, 
that has been a subject of controversy, is achieved via a basic 
version of our 1-bond$\rightarrow$2-mode model originally developed for 
(Zn,Be)-chalcogenides, that exhibit a large contrast in the bond 
properties, and recently extended under a simplified form to 
the usual (Ga,In)As alloy. The Raman/infrared features from (Ga,In)P 
are accordingly re-assigned, with considerable change with respect 
to the previous approaches. In particular the In impurity mode, 
previously assigned within ($\sim$390~cm$^{-1}$) the optical 
band of the host GaP compound (368--403~cm$^{-1}$), is re-assigned 
below it ($\sim$350~cm$^{-1}$). Accordingly the Ga--P and In--P 
transverse optical branches do not overlap, which reconciles 
(Ga,In)P with the Elliott's criterion. Besides, we show that 
the idea of two bond lengths per species in alloys, supported 
by our 1-bond$\rightarrow$2-phonon picture, opens an attractive area for 
the discussion of spontaneous ordering in GaInP$_2$, and in mixed 
crystals in general. Essentially this is because it allows to 
play with the related competition effects regarding the minimization 
of the local strain energy due to the bond length mismatch between 
the parent compounds. In particular the unsuspected issue of 
intrinsic limit to spontaneous ordering comes out ($\eta\sim$0.5
in GaInP$_2$). The whole discussion is supported 
by detailed re-examination of the (Ga,In)P Raman/infrared data 
in the literature, full contour modeling of the transverse and 
longitudinal optical Raman lineshapes via our phenomenological 
1-bond$\rightarrow$2-mode model, and first-principles bond length 
calculations concerned with the minority bond species close to the impurity 
limits (Ga $\sim$ 0, 1) and to the Ga--P (Ga $\sim$0.19) and 
In--P (Ga $\sim$0.81) bond percolation thresholds. In the latter 
case we discriminate between connected and isolated bonds, not 
in the usual terms of next-nearest neighbors.
\end{abstract}
\maketitle

\section{INTRODUCTION}
The long wave vibrational properties of (A,B)C semiconductor 
mixed crystals, where C denotes indifferently the anionic or 
the cationic species, are well-documented both experimentally 
and theoretically. This has lead to a crude classification of 
the phonon mode behavior of \textit{random} mixed crystals in the 
Raman or infrared (IR) spectra in two categories.$^{1}$ Most random 
mixed crystals exhibit the so-called 1-bond$\rightarrow$1-mode behavior. 
This corresponds to well-separated A--C and B--C transverse-longitudinal 
optical (TO--LO) bands over the whole composition range. These 
degenerate into distinct AC:B (B$\sim$0) and BC:A (A$\sim$0) 
impurity modes, and have strengths that scale as the corresponding 
fractions of bonds in the crystal. The remaining random alloys 
exhibit the so-called 2-bond$\rightarrow$1-mode behavior, also referred to 
as the mixed-mode behavior. This corresponds to a single TO--LO 
band with (A--C, B--C)-mixed character. The frequency range that 
the band covers shifts continuously from one end member to the 
other when the alloy composition changes, and the strength remains 
approximately constant throughout the whole composition range. 
In mixed-mode mixed crystals the impurity modes are expected 
to fall within the optical bands of the host lattices, and thereby 
a common believe is that they should not be observed as distinct 
and separate modes.$^{2}$ 

A rather crude criterion is that for 1-bond$\rightarrow$1-mode behavior the 
TO-LO bands of the parent materials must not overlap. As a matter 
of fact the condition is too strong, but if overlap is large, 
the 2-bond$\rightarrow$1-mode behavior is always found. 
Elliott \textit{et al.}$^{3}$ 
derived a more accurate criterion, even though simple, based 
on the Coherent Potential Approximation. This is now the standard 
to decide about the 1-bond$\rightarrow$1-mode or mixed-mode behavior of the 
random mixed crystals. Basically for 1-bond$\rightarrow$1-mode behavior the 
relative change in the reduced mass of the bond induced by the 
impurity must be greater than the difference between the limit 
dielectric constants of the host lattice, normalized to the sum. 
It is worth to notice that this criterion neglects disorder in 
the force constants, only mass disorder is taken into account.

There would be only one fascinating exception that does not fit 
in the above classification of phonon mode behavior, i.e. (Ga,In)P.$^{4}$ 
According to the Elliott's criterion, random (Ga,In)P should exhibit 
a typical 1-bond$\rightarrow$1-mode behavior in the Raman/IR spectra. 
Elliott \textit{et al.} emphasize that the larger the difference between 
the average mass of the substituting species on one side, and the mass on 
the unperturbed site on the other side, the more reliable their criterion. 
Precisely this is true for (Ga,In)P as both In (4$^{th}$ row 
in the periodic table) and Ga (3$^{rd}$ row) are much heavier than 
P (2$^{nd}$ row). As a matter of fact, the TO--LO bands from pure 
InP (303--345~cm$^{-1}$) and pure GaP (368--403~cm$^{-1}$) do not overlap. 
However, the 1-bond$\rightarrow$1-mode behavior does not show up in the 
Raman/IR spectra. 

In the following we propose a brief survey of the extended vibrational 
information available for random (Ga,In)P in the literature, 
and of the different models, for clear insight upon the puzzling 
phonon behavior of this alloy. We proceed chronologically, for 
more clarity. 

Lucovsky \textit{et al.}$^{5}$ have realized the first vibrational study 
of (Ga,In)P alloys, by using a combination of far-IR reflectance 
and Raman scattering on polycrystalline samples. They observed 
a main TO--LO band that behaved as expected in case of a mixed-mode 
behavior (refer above). In addition a minor TO-LO band was evidenced 
within the main TO-LO band. Surprisingly the minor LO mode shows 
up at lower frequency than the TO counterpart, which discouraged 
an assignment of the minor TO-LO band in terms of a zone-center 
signal. This was rather attributed to zone-edge modes. The authors 
concluded that (Ga,In)P was the first example of a mixed-mode 
behavior among {\bf III}--{\bf V}'s. The impurity modes were tentatively 
located at $\sim$330~cm$^{-1}$ (GaP:In) and $\sim$390~cm$^{-1}$ 
(InP:Ga), i.e. within the optical bands of the host lattices.

Beserman \textit{et al.}$^{6}$ have reported an exhaustive Raman study 
of a large amount of polycrystalline (Ga,In)P alloys made of 
small pieces of single crystals. The whole composition range 
could be analyzed. Both the TO and LO modes were allowed in their 
scattering geometries. First, Beserman \textit{et al.} showed that 
the minor TO--LO band persists all the way from the intermediate 
composition range down to both the Ga- and In-dilute limits. 
In addition the authors could perform proper LO symmetry analysis 
by using a convenient piece of single crystal with small In content 
($\sim$0.04). Somewhat surprisingly the minor LO mode, localized 
at $\sim$390~cm$^{-1}$, was found to be highly polarized, just 
as the main zone-center LO mode. The apparent zone-center 2-modes 
LO behavior at small In content was attributed to a splitting 
of the nominal zone-center LO mode due to coupling with two-phonon 
combinations from the zone-edge. Besides, Beserman \textit{et al}. 
derived decisive information in the TO symmetry, which we detail 
below [refer to point (i)].

Jahne \textit{et al.}$^{7}$ discussed the minor TO-LO band as an individual 
zone-center response, which made it difficult to cover the mixed-mode 
behavior. For modeling of their IR spectra obtained with polycrystalline 
(Ga,In)P ingots, Jahne \textit{et al.} used a version of the cluster 
model originally developed by Verleur and Barker to account for 
the complex Raman/IR spectra of Ga(As,P)$^{8}$ and Cd(S,Se)$^{9}$, that 
were attributed to local segregation effects. In this model the 
entire mixed crystal is built up from five basic units corresponding 
to the possible first-neighbor arrangements around the unperturbed 
site. Basically the minor TO--LO band would have its origin in 
the change of the Ga--P force-constant from one type of basic 
unit to another. However, a model with potentially four oscillators 
per bond species in the alloy seems oversized to account for 
a single additional mode in the Raman/IR spectra. 

At this stage further discussion of the phonon mode behavior 
of (Ga,In)P clearly required deeper insight upon the impurity 
modes. Careful Raman investigation of (Ga,In)P samples in the 
Ga- and In-dilute limits was achieved by Jusserand and Slempkes$^{10}$ 
(JS) and Kato \textit{et al}.,$^{11}$ respectively. For this purpose a 
new generation of samples with well-defined symmetry was used, 
i.e. (Ga,In)P samples grown as epitaxial layers or single crystals. 

JS did perform Raman measurements with mostly In-rich (Ga,In)P 
epitaxial layers. They have used a standard backscattering geometry 
along the [001]-growth axis. With this geometry only the LO modes 
are allowed, the TO modes are forbidden. JS did observe the progressive 
emergence of the GaP-like LO line on the high-frequency side 
of the dominant InP-like LO line from very low Ga incorporation 
($\sim$0.1\%). The InP:Ga impurity mode was accordingly identified 
as a distinct mode at $\sim$2~cm$^{-1}$ above the InP LO mode, 
contrary to first expectations. Interestingly, JS noted that 
the In--P and Ga--P lines exhibit similar strengths for as small 
Ga incorporation as 1.5\%. This was attributed to some coupling 
between the two LO modes, due to their vicinity. However, the 
discussion remained qualitative only, and the physical mechanism 
behind was not identified. 

With the InP:Ga impurity mode being out of the InP optical band, 
as for a typical 1-bond$\rightarrow$1-mode system, and the GaP:In impurity 
mode being within the GaP optical band, as for a mixed-mode system, 
a strong overlapping of the Ga-P and In-P bands is expected in 
the alloy, resulting in a complicated phonon behavior. JS$^{10}$ 
proposed that (Ga,In)P is an exceptional alloy in the classification 
of phonon mode behavior. It would obey the so-called modified 
2-mode behavior, with a dominant TO mode at low-frequency that 
joins the InP (303~cm$^{-1}$) and GaP (368~cm$^{-1}$) parent TO modes, 
and a minor TO mode at high-frequency that connects the InP:Ga 
(347~cm$^{-1}$) and GaP:In (390~cm$^{-1}$) impurity modes. 

Kato \textit{et al.}$^{11}$ performed a thorough Raman study of (Ga,In)P 
single crystals and epitaxial layers covering the whole composition 
range, with special emphasis upon the Ga-rich side. They could 
confirm the existence of a quasi-degenerate TO-LO mode 
at $\sim$390~cm$^{-1}$ in the In-dilute limit, which at first sight supported 
the original assignment in terms of the GaP:In impurity mode, and thereby 
the phonon picture proposed by JS. Interestingly, we note from 
their exhaustive data that the TO mode at high-frequency, earlier 
referred to as the minor TO mode, does not seem so `minor'. In 
particular close to the stoichiometry (In$\sim$0.5) the low- 
and high-frequency TO modes have similar strengths (refer to 
fig. 5 in Ref. 11). This has attracted little attention so far.

In the past decade most of the attention was focused on 
Ga$_{0.51}$In$_{0.49}$P, 
abbreviated GaInP$_2$ for simplicity, which has become the leader 
material for the study of spontaneous ordering in semiconductor 
mixed crystals.$^{4}$ Precisely to close this brief overview we 
mention the reference first-principles calculations of the transverse 
phonon density of states (DOS) in disordered GaInP$_{2}$, recently 
performed by Ozolin\v{s} and Zunger (OZ).$^{12,13}$ A theoretical dielectric 
function $\varepsilon$ was derived for pre-insight upon the TO Raman 
lineshapes, via ${\rm Im}(\varepsilon)$. 
What emerged is that the dominant TO mode around 330~cm$^{-1}$ 
involves both Ga--P and In--P vibrations, as expected. Our view 
is that such behavior is consistent with the mixed-mode description 
as well as with the modified 2-mode alternative, as proposed 
by Lucovsky \textit{et al}. and JS, respectively. Also, OZ derived 
decisive insight upon the minor TO mode, which is summarized 
below [refer to point (ii)]. 

Following JS the consensus so far is that random (Ga,In)P exhibits 
the modified 2-mode behavior in the Raman/IR spectra, which 
contradicts the Elliot's criterion. In addition the modified 
2-mode picture fails to explain two key TO features that have 
attracted little attention so far:

(i) 
the dominant TO mode exhibits a marked antagonist asymmetry on 
each side of the stoichiometry (Ga $\sim$0.5), as detected 
by Beserman \textit{et al.} by using Raman scattering (see Fig. 2 
in Ref. 6). This was attributed to a Fano interference with the 
disorder-induced combination of transverse (TA) and longitudinal 
(LA) acoustical modes at the $X$ zone-edge, that was assumed to 
be at higher frequency than the zone-center TO mode in pure InP, 
as opposed to GaP. However, later measurements of the phonon 
dispersion in InP invalidate this mechanism. Indeed the TA($X$)+LA($X$) 
acoustical combination was found below the zone-center TO mode 
in InP (Ref. 14), as in GaP (Ref. 15). 

(ii) 
the minor TO mode in disordered GaInP$_2$ has a GaP-like character, 
as shown by OZ in their first-principles calculations (see Fig. 3-a 
in Ref. 12). This, in particular, is challenging for the 
modified 2-mode picture, as the latter implies a (Ga--P, In--P)-mixed 
character for both the dominant and the minor TO modes, at any 
alloy composition. 

In this work we investigate whether our 1-bond$\rightarrow$2-mode picture, 
earlier referred to as the `percolation' picture, may provide 
consistent understanding of the puzzling phonon behavior of (Ga,In)P, 
which is still lacking. This picture was originally developed 
for the long-wave phonons of (Zn,Be)-chalcogenides,$^{16}$ that 
opened the class of mixed crystals with contrast in the bond 
force constant, and has been extended recently to (Ga,In)As, 
with much success.$^{17}$ 

The key issue when considering the physical properties of random 
(A,B)C mixed crystals is how to handle the problem of alloy disorder. 
Certainly the most convenient way is to describe the system in 
terms of the virtual crystal approximation (VCA). Accordingly 
each atom C from the unperturbed site is ideally surrounded by 
four virtual nearest neighbors, each of these consisting of a 
statistical average of the A and B substituting species depending 
on the alloy composition. This way perfect order is artificially 
re-built in the crystal, where it does not exist in reality, 
so that the approaches finalized at the microscopic scale for 
the basic understanding of the physical properties of the perfectly 
ordered parent compounds can be directly extended to the alloys. 
Regarding vibrational properties, such an approach leads to the 
idea that each bond in the alloy should bring a single feature 
in the Raman/IR spectra, as in the corresponding parent compound, 
but with characteristics (strength, frequency) depending on the 
alloy composition. This corresponds to a typical 1-bond$\rightarrow$1-mode 
behavior in the Raman/IR spectra, as accounted for by the well-admitted 
modified-random-element-isodisplacement (MREI) model developed 
by Chang and Mitra,$^{1}$ based on a VCA description of the mixed 
crystals. Provided some adjustment is made, this model also accounts 
for the mixed-mode behavior.$^{2}$ 

While the VCA seems actually relevant for the integral physical 
properties of semiconductors, that operate a natural average 
on alloy disorder, such as the band gap or the lattice constant, 
our view is that it should not apply to vibrational properties 
because these address directly the \textit{bond force constant}, which 
is a local physical property. We claim that their basic understanding 
requires detailed insight upon the local neighborhood of the 
substituting species, which falls into the scope of the percolation 
site theory.$^{18}$ Essentially this is concerned with the statistical 
properties (population, internal structure, \dots ) of clusters 
formed by sites occupied at random on a regular lattice. While 
there is an obvious analogy between the topology of such systems 
and the topology of the random mixed crystals, we note that the 
concept of percolation remains basically outside the schemes 
used for the very basic understanding of the routine physical 
properties of mixed crystals.

In our 1-bond$\rightarrow$2-mode picture for the Raman/IR spectra of the 
random (A,B)C mixed crystals we describe these as true composite 
media made of two coexisting A-rich and B-rich regions, resulting 
from natural fluctuations in the alloy composition at the local 
scale. This way the alloy disorder is explicitly recognized, 
while it was totally eclipsed with the MREI-VCA description. 
Separate resolvable phonon modes are envisioned for each bond 
species, corresponding to the different force constants experienced 
in each of the A-rich and B-rich regions. In particular our phenomenological 
model envisages singularities in the bond force constant, and 
thereby in the phonon behavior, at the bond percolation thresholds. 
We remind that the bond percolation thresholds are the critical 
compositions corresponding to the first formations of pseudo-infinite 
chains of the B--C and A--C bonds in the A$_{1-x}$B$_x$C crystal. By 
using computer simulation based on random substitution on the 
(A,B) cfc sub-lattice, these were estimated by Stauffer as 
$x_{\rm B-C}\sim$0.19 and $x_{\rm A-C}\sim$0.81, respectively.$^{18}$ 
By crossing the percolation threshold the host region undergoes 
a dispersion$\leftrightarrow$pseudocontinuum 
topological transition. As a matter of fact a singularity at 
the bond percolation threshold was clearly observed in the Raman 
spectra of the reference (Zn,Be)Se (see Fig. 1 in Ref. 16) and 
(Zn,Be)Te$^{19}$ systems. The nature of the singularity is briefly 
discussed in Sec. II. It was tentatively discussed in terms of 
the different internal structures of the dispersion and the pseudocontinuum, 
as predicted by the percolation site theory. Detail is given 
in Ref. 17. It is worth to notice that a similar singularity 
at the bond percolation thresholds was also observed by Bellaiche \textit{et 
al.}$^{20}$ in their first-principles calculations dedicated to another 
local physical property of random mixed crystals, i.e. the \textit{bond 
length}. There again the singularity was discussed in terms of 
a percolation behavior. 

The microscopic mechanism for the 1-bond$\rightarrow$2-mode behavior in the 
Raman/IR spectra was previously identified as the difference 
in bond length due to the different local bond distortions according 
to whether the bonds belong to the randomly formed A-rich or 
B-rich regions.$^{21}$ On this basis the larger the contrast between 
the bond lengths, the more clearly the 1-bond$\rightarrow$2-mode behavior 
is expected to show up in the Raman/IR spectra. In fact the contrast 
is large for (Ga,In)P, as detailed below. At this stage, let 
us emphasize that our 1-bond$\rightarrow$2-phonon picture can not be derived 
from the Elliott's criterion because the latter considers mass 
disorder only, not disorder in the force constant.$^{3}$

Basically we expect two well-separated phonon branches for the 
Ga-based bond and two tight ones for the In-based bond in (Ga,In)P, 
as for (Ga,In)As.$^{17}$ Actually (Ga,In)As can be used as a reference 
for the study of alloying effects on the phonon properties of 
(Ga,In)P, because the substituting species are the same, hence 
a similar contrast in the bond properties and thereby similar 
local bond distortions in the mixed crystals, with concomitant 
impact upon the phonon frequencies. While P has a smaller covalent 
radius (1.06 {\AA}) than As (1.20 {\AA}), so that the P-based bonds 
are shorter/stiffer than the As-based ones, the contrasts between 
the bond lengths ($l$) and the bond stiffness, best described by the ratio 
$R$ between the bond stretching ($\alpha$) and the bond bending ($\beta$) 
force constants, remain similar in (Ga,In)As 
($\Delta l/l\sim$6.3\%, $\Delta R/R\sim$28.1\%) and (Ga,In)P 
($\Delta l/l\sim$7.1\%, $\Delta R/R\sim$34.3\%),$^{22}$ 
the Ga-based bond being shorter/stiffer. Besides, we have checked that 
(Ga,In)As is classified as a 1-bond$\rightarrow$1-mode system according to 
the Elliott's criterion, as (Ga,In)P. 

We discuss the TO modes mainly, because these consist of purely 
mechanical vibrations, i.e. quasi-independent oscillators, and 
thereby carry reliable strength/frequency information on each 
oscillator. We have shown in earlier work that proper investigation 
of the 1-bond$\rightarrow$2-mode behavior via the raw LO Raman data 
is basically hopeless due to strong coupling between neighbor individual LO 
modes via their common long range polarization field $\vec{E}$ .$^{16}$ 
Accordingly our LO study here entirely proceeds from the safe TO modes.

The paper is organized as follows. In Sec. II we remind briefly 
the basis of our phenomenological 1-bond$\rightarrow$2-mode picture for 
contour modeling of the (Ga,In)P Raman lineshapes, and we outline 
the ab initio calculations that we implement to support our corresponding 
re-assignment of the Raman/IR features. In Sec. III our attention 
is focused on the phonon behavior of random (Ga,In)P. In Sec. 
III-A we re-assign the GaP:In impurity mode below the GaP TO--LO 
optical band from careful re-examination of the Raman/IR data 
in the literature, and propose on this basis a simple version 
of our 1-bond$\rightarrow$2-phonon picture that applies to random (Ga,In)P. 
In Sec. III-B we perform ab initio calculations of the bond length 
distribution of the minority bond species in large (Ga,In)P supercells 
corresponding to alloy compositions close to the dilute limits 
and to the In--P (In $\sim$0.19) and Ga--P (In $\sim$0.81) 
bond percolation thresholds. We discriminate between connected 
and isolated bonds, not in the usual terms of next-nearest neighbors. 
The ab initio calculations are used to validate our novel assignment 
of the GaP:In impurity phonon mode, and to provide qualitative 
insight upon the magnitude of the phonon splitting $\delta$ 
within each of the Ga--P and In--P double-branches. In Sec. III-C 
we show that our model provides consistent understanding of the 
whole phonon behavior of random (Ga,In)P, as summarized in Sec. I. 
In Sec. IV we tackle the key issue of spontaneous ordering 
in GaInP$_2$ on this novel basis. We propose a possible mechanism 
that accounts for the disconcerting evolution of the Raman/IR 
lineshapes with increasing ordering. Conclusions are outlined 
in Sec. V.

\section{PHENOMENOLOGICAL MODEL AND FIRST-PRINCIPLES CALCULATIONS}

First, we outline briefly, for more clarity, the technical aspects 
of our phenomenological 1-bond$\rightarrow$2-mode model for long wave phonons 
in random zincblende A$_{1-x}$B$_x$C mixed crystals. Extensive detail 
is given elsewhere.$^{16}$ 

We start with the frequency aspect. Schematically, our view is 
that for each bond species the TO response over the whole composition 
range consists of two quasi-parallel branches tied up to the 
corresponding parent and impurity modes at the two ends of the 
composition range, and separated by a characteristic finite frequency 
gap $\delta$ in the dilute limits. Somewhat surprisingly, for 
a given bond species, the bonds are longer (shorter) within the 
host region that refers to the parent material with the smaller 
(larger) lattice constant. This was explained in detail elsewhere, 
based on first-principles bond length calculations in the two 
host regions.$^{21}$ Now, the shorter the bond length, the larger 
the bond force constant, and thereby the phonon frequency. Accordingly 
within each double-branch the low (high) frequency branch refers 
to the host region with the parent material corresponding to 
the smaller (larger) lattice constant. 

Now we come to the singularity in the bond force constant at 
the bond percolation thresholds, as mentioned in Sec. I. Basically, 
for each individual TO branch the model envisages two different 
regimes on each side of the bond percolation threshold to which 
the host region refers: one in which the optical mode of the 
most dilute substitutional species vibrates with a frequency 
that is basically independent on its concentration (regime 1), 
and one in which the frequency of the same mode depends smoothly 
on the alloy composition according to a traditional but `rescaled' 
modified-random-element-isodisplacement (MREI) description (regime 2). 

We turn next to the strength aspect. A very general trend is 
that the oscillator strengths and Faust-Henry coefficients from 
the overall A--C and B--C signals in the Raman/IR spectra do scale 
as the fraction of the related oscillators in the A$_{1-x}$B$_x$C 
alloy, i.e. as $(1-x)$ and $x$, respectively.$^{1}$ These are referred 
to as the global weighting factors. Now, simple symmetry considerations 
guarantee that in case of random substitution the scattering 
volumes from the A-rich and B-rich regions scale as $(1-x)$ and 
$x$, respectively. These are referred to as the individual weighting 
factors. They fix the sharing of the available oscillator strength 
and Faust-Henry coefficient within each double-branch. Eventually 
the multiplication of the global and individual weighting factors 
determines the relative strength of the four TO modes (2 AC-like 
and 2 BC-like) in the Raman/IR spectra. 

Once the two double-branches are properly set, full contour modeling 
of the multi-phonon Raman lineshapes is eventually achieved on 
the above frequency/strength basis, while using no adjustable 
parameter, by using the generic equation 
\begin{equation}
I\propto {\rm Im} \left\{ -\varepsilon _{r}^{-1} \left[
1+\sum\limits_{p}\!C_{p}\,K_{p}\,L'_{p}  \right] ^{2}
+\sum\limits_{p}\!C_{p}^{2}\,\frac{K_{p}^{2}\,L'_{p} }{4\pi Z_{p}^{2} } 
\right\} .
\end{equation}

The TO modes are obtained from the imaginary part of the second 
term, while the full expression provides the LO modes. Here, 
the summation runs over the relevant number of oscillators, that 
may be smaller than four, depending on the alloy (see Sec. III). 

$C_{p}$, $K_{p}$ and $L'_{p}$ are defined on a per-oscillator basis and are, 
respectively, the Faust-Henry coefficient of the $p$-mode, 
its TO frequency squared, and its related Lorentzian responses. 
$Z_{p}$ is defined according to the standard MREI notations;$^{1}$ it relates 
to the oscillator strength $S_{p}$, and thereby to the 
(TO-LO)$_p$-splitting. $S_{p}$ and $C_{p}$ are normalized 
to the fraction of $p$-oscillators in the crystal with respect 
to the corresponding parent values. 
$\varepsilon_{r}$ is the relative dielectric function of the mixed crystal in 
a form generalized to multi-oscillators, as established according 
to the standard MREI scheme.

In our model, once the TO phonon double-branches of the Ga--P 
and In--P bonds are properly set, the Raman/IR lineshapes are 
entirely determined. Each double-branch is fixed by two parameters 
only on top of the frequency of the parent TO mode, i.e., the 
magnitude of the phonon splitting $\delta$, and the frequency 
of the impurity mode. For clear insight upon these two key parameters 
in (Ga,In)P we determine a representative distribution of bond 
lengths, that we have done in a first-principles supercell calculation. 
We applied the calculation method and the computer code SIESTA,$^{23,24}$ 
and allowed full relaxation (of lattice parameters along with 
internal coordinates) in a prototype `percolation-threshold' supercell 
similar to that used in Ref. 21, along with `dilute limit' supercells. 
We used the local density approximation throughout. If comparing 
with previous calculations for the (Ga,In)P system, by OZ$^{12,13}$, 
we used the same type of norm conserving pseudopotentials (constructed 
along the Troullier-Martins scheme, see Ref. 25), as OZ, specifically 
using the following configurations and pseudoization radii 
(in brackets, in Bohr): Ga 4$s^1$(2.00) 4$p^1$(2.00) 3$d^{10}$(1.78), 
In 5$s^1$(2.19) 5$p^2$(2.48) 4$d^{10}$(1.68), 
P 3$s^2$(1.83) 3$p^3$(1.83) 3$d^0$(1.83). However, 
we used a different basis set (localized atom-centered functions, 
vs. planewaves as by OZ), and larger supercells (of 64 atoms, 
vs. a maximum of 16 by OZ). Generally, care must be taken that 
our bond length calculations are strictly valid only at zero 
temperature, which might generate slight discrepancy with respect 
to room temperature, of present interest here.

\section{ONE-BOND$\rightarrow$TWO-PHONON PICTURE FOR RANDOM 
(G\lowercase{a},I\lowercase{n})P}
According to the general criterion of localization derived by 
Anderson,$^{26}$ the condition for clear observation 
of the 1-bond$\rightarrow$2-phonon behavior in the Raman/IR spectra 
of random (Ga,In)P is that the 
Ga-rich and In-rich environments generate fluctuations in the 
TO frequency that are typically larger than the reference TO 
dispersion in the parent material. Precisely the TO mode in GaP 
is nearly dispersionless, with a difference between the frequencies 
of the zone-center and zone-edge ($X$) TO modes of $\sim$1.5~cm$^{-1}$ 
(Ref. 15). This brings a much favorable context for the observation 
of a 1-bond$\rightarrow$2-phonon behavior in the Ga-P spectral range. In 
contrast the same frequency difference is rather large in InP, 
of $\sim$20~cm$^{-1}$ (Ref. 14). Actual phonon localization in 
the In--P spectral range would require extremely large fluctuations 
in the TO frequency, and thereby in the In-P bond length. However, 
by analogy with (Ga,In)As$^{17}$ this is rather unlikely. 

On the above basis we are lead to describe random (Ga,In)P as 
a three-oscillator [1(In--P), 2(Ga--P)] phonon system, in a first 
approximation.

\subsection{TO picture}
We propose the 1-bond$\rightarrow$2-mode TO picture outlined in Fig. 1 
for the random Ga$_x$In$_{1-x}$P alloy (thick lines), from careful 
re-examination of the available Raman/IR data in the literature; 
details are given below. It consists of three quasi-parallel branches: 
two well-separated Ga--P branches that converge in the dilute 
limits, above two In--P branches so tight (dashed lines) that 
they merge into an overall In--P branch. The high and low-frequency 
Ga-P branches refer to Ga--P vibrations within the In-rich and 
Ga-rich regions, respectively (refer to Sec. II). The TO modes 
are accordingly labeled as 
TO$_{\rm In-P}$, TO$_{\rm Ga-P}^{\rm Ga}$ and TO$_{\rm Ga-P}^{\rm In}$
with increasing frequency, where the superscript refers to the 
host region. For each individual Ga--P branch the so-called regime 2 
(refer to Sec. II) is ideally modeled as a straight line, in 
the first approximation. We have checked that the proper re-scaled 
MREI curves exhibit a bowing of merely $\sim$1~cm$^{-1}$. Regarding 
the strength aspect, the TO$_{\rm In-P}$, TO$_{\rm Ga-P}^{\rm Ga}$
and TO$_{\rm Ga-P}^{\rm In}$
modes scale as the corresponding fractions of bonds in the alloy, 
i.e. as $x$, $x^2$ and $(1-x)x$, respectively (refer to Sec. II). 
The relative strengths of the different TO modes are explicitly 
indicated in Fig.1, for more clarity. The Ga--P and In--P TO double-branches 
do not overlap in our description, which reconciles (Ga,In)P 
with the Elliott's criterion.

\begin{figure}[b]
\centerline{\includegraphics[height=0.9\textwidth,angle=270]{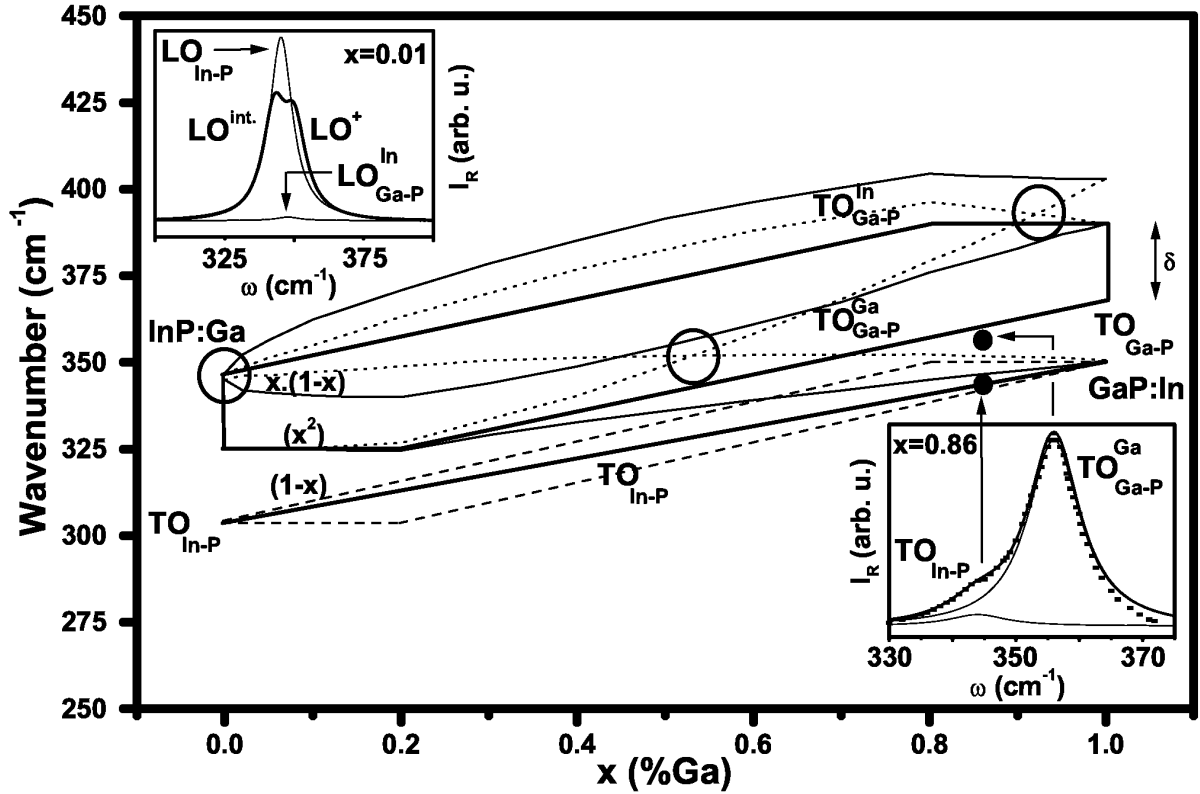}}
\caption{%
Percolation picture for random Ga$_x$In$_{1-x}$P. Thick lines 
refer to TO modes. The corresponding fractions of bonds are indicated 
on the left. Thin and dotted lines refer to coupled and uncoupled 
LO modes, respectively. The circles mark strong $\vec{E}$-coupling regimes. 
Calculated Raman lineshapes in the Ga-dilute limit (LO, top inset) 
and at large Ga-content (TO, bottom inset) are shown. In the latter case 
the experimental data from Kato \textit{et al.} (reproduced from Fig. 6 
of Ref. 11) are added (squares), for comparison.
}
\end{figure}

From Fig. 1 we re-assign the s-called minor and dominant TO modes 
in the Raman/IR spectra as the TO$_{\rm Ga-P}^{\rm In}$ mode and the sum 
of the nearby TO$_{I\rm n-P}$ and TO$_{\rm Ga-P}^{\rm Ga}$ modes, respectively.
Accordingly the minor TO mode is GaP-like, 
as predicted by the first-principles calculations of OZ$^{12,13}$ 
[refer to point (ii) above]. Also, the strengths of the nearby 
TO$_{\rm In-P}$ and TO$_{\rm Ga-P}^{\rm Ga}$ modes vary in opposite sense 
versus the Ga-content $x$, which accounts for the antagonist asymmetry 
of the sum at large and small $x$ values as detected by 
Beserman \textit{et al.} by using Raman spectroscopy 
[refer to point (i) above]. More quantitative insight 
is given in Sec. III-C.

On top of the parent TO--LO bands three empirical parameters were 
used to build up the entire 1-bond$\rightarrow$2-mode picture: the frequencies 
of the GaP:In and InP:Ga impurity modes, as for the more simple 
1-bond$\rightarrow$1-mode MREI picture,$^{1}$ plus the Ga--P splitting, noted 
$\delta$. 

We have taken $\delta\sim$22~cm$^{-1}$ corresponding to the frequency gap 
between the dominant (368~cm$^{-1}$) and the minor (390~cm$^{-1}$) TO modes 
close to the In-dilute limit, as accurately measured by Kato 
\textit{et al.} by using Raman spectroscopy.$^{11}$ This fixes directly the 
profile of the whole TO$_{\rm Ga-P}^{\rm In}$ branch once the singularity 
at the In--P bond percolation threshold is taken into account 
(refer to Sec. II). The TO$_{\rm Ga-P}^{\rm Ga}$
branch is derived by symmetry, as expected in case of a random 
Ga substitution to In over the whole composition range.$^{16}$ On 
this basis a typical phonon splitting of $\sim$33~cm$^{-1}$ should 
be observed between the two Ga--P TO modes at the stoichiometry, 
in remarkable agreement with the value of $\sim$35~cm$^{-1}$ 
found by OZ via first-principles calculations (refer to Figs. 
3 from Refs. 12-13). 

The location of the InP:Ga impurity mode at $\sim$347~cm$^{-1}$, 
as detected by JS$^{10}$ by using Raman spectroscopy, is not questioned 
at this stage, owing to the strong experimental support. On the 
other hand we re-assign the GaP:In mode below the GaP optical 
band, i.e. at $\sim$350~cm$^{-1}$. This is estimated from linear 
extrapolation of the TO$_{\rm In-P}$ frequency from the bulk 
(303~cm$^{-1}$) to the In-dilute limit, 
passing through an intermediary value (dot in Fig. 1) measured 
close to the In-dilute limit for more accuracy. This is adjusted 
so as to achieve full contour modeling of the experimental 
(TO$_{\rm In-P}$ + TO$_{\rm Ga-P}^{\rm Ga}$) Raman signal as obtained 
by Kato \textit{et al.} at the In content 
of $\sim$0.14 (refer to the bottom spectrum from Fig. 6 of Ref. 11), 
the other parameters being otherwise fixed by the 
1-bond$\rightarrow$2-mode model. Technical detail is given in Sec. III-C. 
The same phonon damping was taken for the two modes. The best fit, obtained 
for $\omega$(In--P)=344~cm$^{-1}$, is superimposed to the experimental curve 
in the bottom inset of Fig. 1, for comparison. 

\subsection{Validation via first-principles calculations }
Clear theoretical insight upon the phonon behavior of random 
(Ga,In)P was derived by OZ$^{12,13}$ at the representative alloy 
composition corresponding to the stoichiometry, which was much 
helpful to build up our TO phonon picture displayed in Fig. 1 
[refer to point (ii) above]. Additional phonon calculations at 
other alloy compositions are of little interest for our purpose 
because the relative positions of the three TO modes is not expected 
to be much dependent on the alloy composition, as can be inferred 
from Fig. 1. What we rather need to fully validate Fig. 1 is 
complementary bond length information close to other critical 
alloy compositions, i.e. the bond percolation thresholds and 
the dilute limits, as detailed below. We address two issues.

First, we need direct evidence that the GaP:In impurity mode 
stays below the GaP optical band, not within (390~cm$^{-1}$) as 
is currently admitted. For direct insight we achieve full relaxation 
of a (Ga,In)P supercell containing one In atom only out of 32 
cations (In $\sim$3 at.\%). The single-site In substitution 
to Ga shortens the In--P bond length from 
$l_{0}\sim$2.540~{\AA} in the pure InP crystal to 
$l_{\rm imp.}\sim$2.485~{\AA} because the In impurity has to fit into the 
GaP-like host media characterized by a smaller bond length. The 
key point is that the local In--P compression is hydrostatic here. 
Accordingly, while explicit ab initio calculations of the phonons 
either by the linear response, as was done by OZ,$^{12,13}$ or by 
the frozen phonon scheme, as in Ref. 21, are the obliged way 
to get theoretical phonon insight out of the dilute limits, where 
the individual bonds undergo complex bond distortions, they are 
not required here in a first approximation. Instead we simply 
estimate the expected shift $\Delta\omega _{\rm T}^{2}$
in the square TO frequency due to the local compression via 
the Gr\"{u}neisen parameter 
$\gamma_{\rm T}\sim$1.44$\pm$0.02, as measured from the frequency-dependence 
of the zone-center TO mode of pure InP under hydrostatic pressure.$^{27}$ 
We use the relation$^{28}$ 
\begin{equation}
\frac{\Delta \omega_{\rm T}^2}{\omega_{\rm T}^2} = -6\gamma_{\rm T} \cdot
\frac{\Delta l}{l}\,,
\end{equation}
where $\Delta l$ is the relative change in the bond length. This yields 
the estimate of $\omega\sim$330$\pm$0.6~cm$^{-1}$ for the displaced 
frequency of the GaP:In impurity mode, i.e. even lower than the 
measured value ($\sim$350~cm$^{-1}$).

We have performed similar bond length calculations by using the 
same supercell but with Ga and In interchanged. 
$\gamma_{\rm T}\sim$1.09$\pm$0.03 for pure GaP$^{27}$ so that the observed 
lengthening of the Ga--P bonds from $l_{0}\sim$2.360~{\AA} in pure GaP to 
$l_{\rm imp.}\sim$2.393~{\AA}, due to outward hydrostatic relaxation of the 
four P neighbors around the Ga impurity, yields the expected 
frequency $\omega\sim$350$\pm$0.05~cm$^{-1}$ for the InP:Ga 
impurity mode, in very good agreement with the actually measured 
value ($\sim$347~cm$^{-1}$, see Ref. 10). This validates our 
procedure to estimate the impurity-related phonon frequencies.

As the second issue we need evidence of a 1-bond$\rightarrow$2-mode bond 
length distribution, that would mirror the 1-bond$\rightarrow$2-mode phonon 
behavior. Basically the bond length distribution must be narrow 
for the In-P bond, and well-resolved for the Ga-P one. Also, 
the short (long) bonds must refer to the In-rich (Ga-rich) region, 
for each species. To address these aspects we consider the minority 
bond species close to the Ga--P (In$\sim$0.81) and In--P 
(In$\sim$0.19) bond percolation thresholds, where it is easy to figure out 
the Ga-rich and In-rich regions. We use our prototype (Zn,Be)Se supercell 
of Ref. 21, with 4 impurity atoms connected in a straight wall-to-wall 
chain plus 2 isolated impurities. The Ga and In atoms are interchanged 
for the In-poor and Ga-poor configurations. In the Ga-poor supercell 
the connected and isolated Ga-P bonds refer to the Ga-rich and 
In-rich regions, respectively. The situation is reversed for 
the In-P bonds in the In-poor supercell.

First, we consider the Ga-poor supercell. We obtain an overall 
bi-modal bond length distribution for the stiff-short Ga-P bond 
(not shown) that mimics the Be-Se reference in (Zn,Be)Se (refer 
to Fig. 4 top-left in Ref. 21), as expected. Only, the distributions 
of the isolated and connected bonds are re-centered close to 
the bond length of pure GaP, i.e. at $\sim$2.383~{\AA} and $\sim$2.402~{\AA}, 
respectively. The bond length difference is $\sim$8\textperthousand, 
which has to be related to $\delta\sim$22~cm$^{-1}$ on the phonon side, 
as detailed above. Remarkably, $\delta\sim$0 for the Ga--As bond in (Ga,In)As, 
while the difference in the Ga--As bond length reaches $\sim$1\%.$^{17}$ 
This is consistent with our present view that for a given local bond distortion
$\delta$ increases with the bond stiffness. This statement is based on 
our earlier observation in the reference (Zn,Be)VI systems that 
similar changes in the Be--{\bf VI} ($\sim$2\%) and Zn--{\bf VI} ($\sim$1\%) 
bond lengths according to whether the bonds belong to the Be-rich 
region or the In-rich one generate a much larger phonon shift, 
i.e. a much larger change in the bond force constant, for the 
stiff Be--{\bf VI} bond ($\sim$8.5\% of the TO frequency in pure 
BeVI) than for the soft Zn--{\bf VI} one ($\sim$1\% of the TO frequency 
in pure Zn{\bf VI}).$^{17}$ On this basis we anticipate a large 
$\delta $ value for the Ga-N bond in zincblende (Ga,In)N, with larger 
contrast in the bond properties as (Ga,In)As and (Ga,In)P, i.e. 
($\Delta l/l \sim$9.5\%, $\Delta R/R \sim$41.7\%). This is an unexplored issue.

In the In-poor supercell the connected In--P bonds are shorter 
(2.487~{\AA}) than the isolated (2.493~{\AA}) ones, as expected. Also, 
the difference in bond length is small, i.e. $\sim$2{\textperthousand}, 
as for the reference In--As bond in (Ga,In)As.$^{17}$ We expect 
that the corresponding small phonon splitting is screened by 
the large TO dispersion in InP ($\sim$20~cm$^{-1}$), and/or by 
the phonon damping. This justifies a posteriori our approximation 
of an apparent 1-bond$\rightarrow$1-mode behavior in the Raman response 
of the In-P bond in Fig. 1.

\subsection{Raman lineshapes of random (G\lowercase{a},I\lowercase{n})P}
Full contour modeling of the Raman lineshapes of random (Ga,In)P 
is achieved by using the reduced (TO) and extended (LO) forms 
of Eq. (1) from the frequency/strength information displayed 
in Fig. 1. The other input parameters are the Faust-Henry coefficients 
of GaP ($-$0.53, Ref. 29) and InP ($-$0.46, Ref. 30) and the parent 
oscillator strengths, as derived according to the traditional MREI scheme 
from the parent TO-LO optical bands and the $\varepsilon_{\infty}$ 
values of GaP (8.45, Ref. 2) and InP (9.53, Ref. 30). Now we 
compare the raw theoretical lineshapes with the Raman data and 
first-principles phonons calculations available in the literature. 
We emphasize that no adjustable parameter is used in our phenomenological 
calculations.

The theoretical TO Raman lineshapes are shown in Fig. 2. A small 
phonon damping of 1~cm$^{-1}$ is taken for a clear overview of the 
whole collection of individual modes. In particular Fig. 2 provides 
straightforward insight upon the puzzling antagonist asymmetry 
of the dominant (TO$_{\rm In-P}$ + TO$_{\rm Ga-P}^{\rm Ga}$) mode 
on each side of the stoichiometry, as detected by Beserman 
\textit{et al.} [refer to point (i) above]. Remarkably the nearby 
TO$_{\rm In-P}$ and TO$_{\rm Ga-P}^{\rm Ga}$ modes have similar strengths 
at the stoichiometry, as ideally expected. The difference is less than 10\% 
to the advantage of the GaP-like mode, in close agreement with the zone-center
TO density of states derived by OZ in their first-principles calculations 
(see Fig. 3-a in Ref. 12). We note that the relative strength 
of their TO mode at $\sim$370~cm$^{-1}$ is small with respect 
to our prediction, apparently due to overdamping. Otherwise the 
variation of our theoretical strength ratio between the 
TO$_{\rm Ga-P}^{\rm In}$ mode (minor) and 
the (TO$_{\rm In-P}$ + TO$_{\rm Ga-P}^{\rm Ga}$) mode (dominant) versus 
the Ga-content $x$ fairly reproduces the 
experimental curve obtained by Kato \textit{et al.} from their exhaustive 
Raman data (see Fig. 5 in Ref. 11), as shown in the inset of 
Fig. 2. In particular the different slopes observed in the dilute 
limits are rather well-reproduced by our model. To the best of 
our knowledge this is the first attempt to put a model on these 
data.

\begin{figure}[b]
\centerline{\includegraphics[height=0.9\textwidth,angle=270]{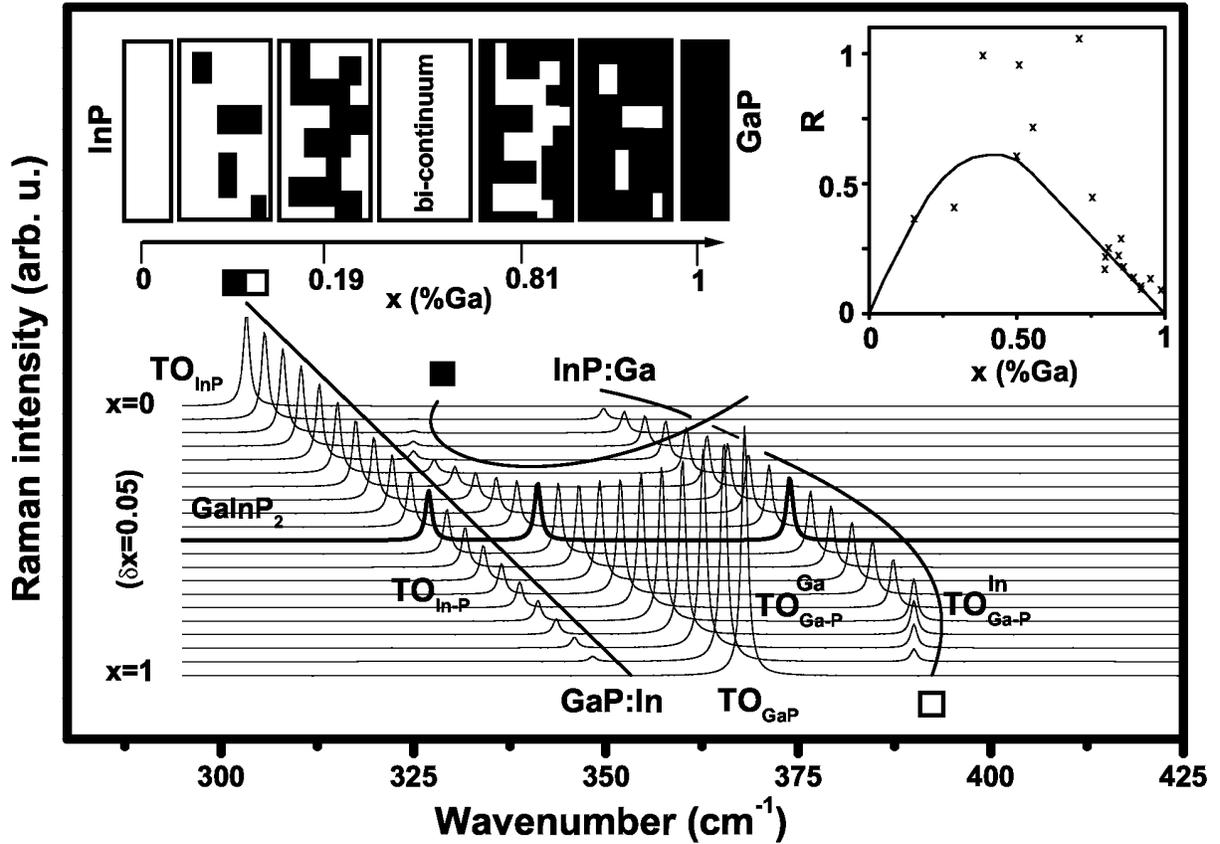}}
\caption{
TO Raman lineshapes of Ga$_x$In$_{1-x}$P calculated from Fig. 1. 
The topologies of the Ga-rich (black) and In-rich (white) 
regions are schematically represented at different $x$-values. 
The modes are accordingly labeled for clarity. The evolution 
of the theoretical strength ratio $R$ between the TO$_{\rm Ga-P}^{\rm In}$
mode and the (TO$_{\rm In-P}$ + TO$_{\rm Ga-P}^{\rm Ga}$) mode versus $x$ 
is shown in the inset. The experimental data 
from Kato \textit{et al.} (taken from Fig. 5 of Ref. 11) are added 
(crosses), for comparison. 
}
\end{figure}

Now we discuss briefly the LO modes. 
$\vec{E}$-coupling is explicitly allowed in our phenomenological model 
by taking a single dielectric function for the three-oscillator 
system. This tends to generate a single `giant' oscillation that 
receives most of the available oscillator strength, thereby blue-shifted 
from the rest of the series.$^{16}$ Three LO features are obtained 
but these have (In--P, Ga--P)-mixed character, which results in 
a strong distortion of the overall LO signal with respect to 
the reference uncoupled LO lines. Examples are given in sec. 
IV. The mixed LO features are simply labeled as 
LO$^{-}$, LO$^{\rm int.}$ and LO$^{+}$ with increasing frequency, 
where superscript `int.' stands for 
`intermediary'. The uncoupled LO lines are calculated via Eq. 
(1) also, but by taking separate dielectric functions for each 
oscillator, i.e. by treating each oscillator independently. They 
are referred to as LO$_{\rm In-P}$, LO$_{\rm Ga-P}^{\rm Ga}$ and 
LO$_{\rm Ga-P}^{\rm In}$ hereafter, by analogy with the TO counterparts. 
The variations of the frequencies of the coupled and uncoupled LO modes versus 
the Ga-content x are displayed as thin and dotted lines in Fig. 1, respectively.

In particular the LO$^{-}$ mode is systematically weak, and remains confined 
between the TO$_{\rm In-P}$ and TO$_{\rm Ga-P}^{\rm Ga}$
phonon branches when the alloy composition changes. In analyzing 
the LO-allowed Raman spectra obtained with the traditional backscattering 
geometry along the [001]-growth axis of the now available (Ga,In)P 
epitaxial layers, we suspect that this mode was previously mistaken 
as the parasitical activation of the dominant TO mode (theoretically 
forbidden), the result of breaking in the Raman selection rules 
induced by the alloy disorder. 

Also, we identify the minor optical band detected by Lucovsky \emph{et al.} 
in their pioneer IR measurements (refer to Sec. I), as the 
LO$^{\rm int.}$ -- TO$_{\rm Ga-P}^{\rm Ga}$
 band. The key point here is that the TO and LO features do not 
refer to the same vibration, which suppresses the enigma of the 
apparent LO-TO inversion.

Three regimes corresponding to strong 
$\vec{E}$-coupling, i.e. quasi-resonance for some of the uncoupled LO 
modes, are identified by circles in Fig. 1. In each case typical 
coupling-induced anti-crossing behaviors are observed (compare 
the thin lines that cross each other, and the dotted lines that 
do not cross in Fig. 1). In particular, strong 
$\vec{E}$-coupling occurs close to the stoichiometry, which prevents the 
discussion of any of the LO features in the Raman/IR spectra 
in this very sensitive composition range (refer to Sec. IV) as 
due to any specific bond vibration. Strong 
$\vec{E}$-coupling occurs also in the Ga-dilute limit, which we identify 
as the mechanism behind the spectacular emergence of the InP:Ga 
mode in the LO Raman spectra, as predicted by JS.$^{10}$ A priori 
this strong distortion with respect to the uncoupled LO features 
(refer to the thin lines in the top inset of Fig. 1) might have 
generated misleading phonon shifts, with concomitant impact on 
the reliability of the frequency of the InP:Ga impurity mode 
as determined from the raw LO Raman spectra. As a matter of fact 
the exact location of the InP:Ga mode was debated at a certain 
time. To be quite sure we have calculated the LO Raman lineshape 
for a typical Ga content of $\sim$1\% for the two proposed 
locations of the InP:Ga impurity mode, i.e. $\sim$2~cm$^{-1}$ 
above$^{10}$ and below$^{7}$ the LO mode of pure InP ($\sim$345~cm$^{-1}$). 
The two theoretical LO Raman lineshapes exhibit antagonist asymmetries 
(not shown), which indicates that the overall LO signal is extremely 
sensitive to the location of the InP:Ga impurity mode. As a matter 
of fact only the assignment proposed by JS fairly reproduces 
the low-temperature LO Raman spectrum that these authors obtained, 
after the latter spectrum was slightly red-shifted to simulate 
ambient conditions (see spectrum c in Fig. 2 of Ref. 10). 

\section{SPONTANEOUS ORDERING IN G\lowercase{a}I\lowercase{n}P$_2$}
Since the pioneer extended X-ray absorption fine structure (EXAFS) 
obtained by Mikkelsen and Boyce with the representative (Ga,In)As 
system,$^{31}$ it is rather well-admitted that while the lattice 
constant exhibits a quasi-linear variation from one end member 
to the other when the alloy composition varies, each bond species 
tends more or less to keep its natural bond length in the alloy, 
as determined in the pure crystal. Transferred to lattice dynamics 
this simple picture would support the standard 1-bond$\rightarrow$1-mode 
behavior, as envisaged by the MREI model.$^{1}$ However, such a 
simple bond length description offers little flexibility for 
the discussion of spontaneous ordering in mixed crystals. As 
a matter of fact the mechanism behind spontaneous ordering still 
remains a debated issue.$^{4}$

Generally our 1-bond$\rightarrow$2-mode picture for the phonon behavior in 
the Raman/IR spectra of mixed crystals opens an attractive area 
for the discussion of spontaneous ordering, because it brings 
the idea of two bond lengths per species in the alloy. This allows 
to play with the related competition effects regarding the minimization 
of the local strain energy due to the bond length mismatch between 
the end compounds. 

As a matter of fact our recent atomistic calculations of the 
bond length distribution in large (Ga,In)As supercells that mimic 
real random alloys clearly confirm the bi-modal character of 
the bond length distribution related to the short Ga-based bond, 
corresponding to a clear bi-modal behavior in the Raman/IR spectra.$^{17}$ 
The configurations were analyzed to distinguish between bonds from 
the same species that are either inter-connected or isolated, 
not in the usual terms of next-nearest neighbors. The analogy 
is straightforward with (Ga,In)P. There again the short Ga--P 
bond exhibits a clear bi-modal behavior in the Raman/IR spectra 
(refer to Fig. 1), corresponding to a bi-modal bond length distribution 
that discriminates between the `short' and `long' Ga--P bonds from 
the In-rich and Ga-rich regions, respectively (refer to Sec. 
III-B). As the phonon splitting between the two Ga--P TO modes 
remains basically unchanged throughout the whole composition 
range ($\sim$33~cm$^{-1}$, refer to Fig. 1), we expect the same 
for the corresponding difference in the Ga--P bond lengths. This 
should remain close to the value found at the Ga--P bond percolation 
threshold, i.e. $\sim$8{\textperthousand} (refer to Sec. III-B).

Now we focus our attention on GaInP$_2$. At this critical composition 
the Ga--P and In--P bond species are in similar proportion in the 
alloy, and also the two series of Ga--P bonds (corresponding to 
the two Ga--P TO modes in the Raman/IR spectra, refer to Sec. 
II), which enhances the local strain energy. Now, the two series 
of Ga--P bonds undergo a local tensile strain, due to the longer 
In--P bonds. Our view is that spontaneous ordering occurs so 
as to favor a single Ga--P bond length in the alloy, i.e. the 
larger one so as to minimize the local strain energy. Basically 
with increasing order we expect that the topology of the (Ga,In) 
substituting species becomes more like that in the Ga-rich region, 
and thus that the TO$_{\rm Ga-P}^{\rm Ga}$ mode reinforces at the cost of the 
TO$_{\rm Ga-P}^{\rm In}$ mode.

As a matter of fact we note that the minor TO resonance 
(TO$_{\rm Ga-P}^{\rm In}$) is deep in the IR transmission spectra obtained 
by Alsina \textit{et al.}$^{32}$ with nominally random GaInP$_2$, while it is 
shallow for strongly ordered GaInP$_2$ (compare Figs. 1 and 2 in Ref. 32), 
suggesting that the minor TO mode weakens with increasing order. 
OZ$^{13}$ arrive at the same conclusion via first-principles calculations. 
Besides Mestres \textit{et al.}$^{33}$ observe by using Raman scattering 
that the dominant TO mode (TO$_{\rm In-P}$ + TO$_{\rm Ga-P}^{\rm Ga}$) 
strengthens and sharpens with increasing order, as can be expected 
from reinforcement of the GaP-like character.

We denote as $\eta'$ (0$\leq\eta'\leq$1) the fraction of `short' Ga-P bonds 
that has turned `long' due to spontaneous ordering. Now we calculate 
the TO and LO Raman lineshapes while increasing progressively 
$\eta'$ until we obtain fair agreement with the extended data in the 
literature related to the now available GaInP$_2$ films. Care 
must be taken that all of these exhibit spontaneous ordering 
to some extent, as characterized by values of the order parameter 
$\eta$ in the range $\sim$0.1--0.5.$^{34}$ We remind that 
$\eta$ measures the average deviation with respect to equal representation 
of the substituting species in the (111) cationic planes, corresponding 
to the formation of 
a (Ga$_{1+\eta}$In$_{1-\eta}$P$_2$)/(Ga$_{1-\eta}$In$_{1+\eta}$P$_2$) 
[111] superlattice. $\eta$=0 corresponds to the perfectly random situation, 
while $\eta$=1 corresponds to the perfectly ordered CuPt-type GaInP$_2$, i.e. 
a succession of full-Ga and full-In cationic planes along the 
[111] direction. 

For direct comparison with the data we take a realistic phonon 
damping of 10~cm$^{-1}$, and a reduced splitting of $\sim$4~cm$^{-1}$ 
between the TO$_{\rm In-P}$ and TO$_{\rm Ga-P}^{\rm Ga}$
modes as given by the ab initio calculations of OZ (see Fig. 3 in Ref. 12), 
i.e. roughly half our prediction. The central frequency remains the same. 
Incidentally this discrepancy of a few cm$^{-1}$ at the stoichiometry between 
the ab initio calculations and our phenomenological model is rather small 
when considering that the latter model is entirely built up from a reduced set 
of three input parameters taken in the dilute limits (the $\delta$ 
value for the Ga-P phonon splitting, and the frequencies of the 
GaP:In and InP:Ga impurity modes). 

The (TO,LO) situation for random GaInP$_2$ ($\eta=\eta'=0$) is shown 
in Fig. 3-a, for reference purpose. The best agreement 
between the model and the data in the literature is obtained 
for $\eta'\sim$0.7, typically. This corresponds to a dominant TO mode 
at $\sim$332~cm$^{-1}$, a weak TO mode at $\sim$372~cm$^{-1}$, 
and three clear LO signals at $\sim$333, $\sim$362 and $\sim$381~cm$^{-1}$. 
The curves are shown in Fig. 3-b. We note that as $\eta'$
increases the LO$^{\rm int.}$ and LO$^{+}$ modes converge while the 
LO$^{\rm int.}$/LO$^{+}$ strength ratio enlarges. Basically this is enough 
to explain the puzzling reduction of the so-called valley-to-depth ratio 
b/a, as schematically indicated in Fig. 3-b, with increasing 
order.$^{4}$ 

\begin{figure}[b]
\centerline{\includegraphics[height=0.9\textwidth,angle=270]{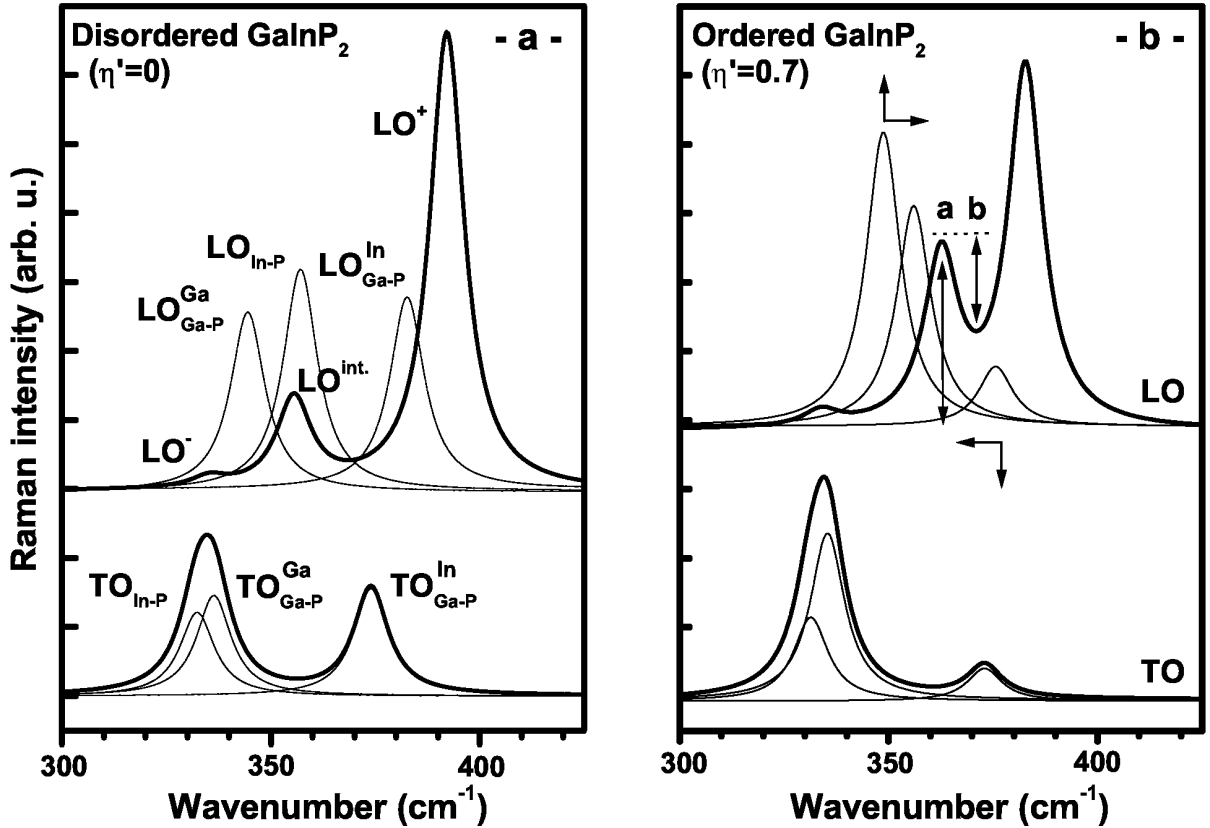}}
\caption{%
Calculated (TO,LO) Raman lineshapes (thick lines) for 
disordered (a, $\eta'$=0) and spontaneously ordered (b, $\eta'\sim$0.7) 
GaInP$_2$. The individual (TO,LO) modes (thin lines) are added, 
for reference purpose. In part (b) arrows indicate 
the strength/frequency variations of the individual GaP-like 
LO modes when $\eta'$ increases. The LO curves are translated 
along the vertical axis, for more clarity. 
}
\end{figure}

At last, we expect an intrinsic limit to spontaneous ordering 
in mixed crystals. In GaInP$_2$ this should be reached when all 
the Ga-P bonds are eventually `long' in the crystal ($\eta'\sim$1). 
At this limit the TO$_{\rm Ga-P}^{\rm In}$ mode, that represents the `short' 
Ga-P bonds in the crystal, should be hardly detectable in the Raman/IR spectra.
Experimentally, this corresponds to $\eta\sim$0.5 (see Fig. 2 in Ref. 35). 
In fact we are not aware that spontaneously ordered GaInP$_2$ films could be 
grown with an order parameter $\eta$ greater than $\sim$0.5. 

\section{CONCLUSION}
The consensus so far was that random (Ga,In)P is the only alloy 
that exhibits the so-called modified 2-mode behavior in the 
Raman/IR spectra, with a dominant TO mode at low frequency that 
joins the parent TO modes and a minor TO mode at high frequency 
that connects the impurity modes. Here we show that (Ga,In)P 
is not an exception in the crude classification of phonon mode 
behavior as established by Elliott \textit{et al.}, to distinguish 
between 1-bond$\rightarrow$1-mode and 2-bond$\rightarrow$1-mode systems. 
Consistent understanding of the phonon mode behavior of (Ga,In)P is achieved 
via a basic version of our 1-bond$\rightarrow$2-mode phenomenological model 
(earlier referred to as the percolation model), supported by 
detailed re-examination of the Raman/IR data available in the 
literature, phenomenological full contour modeling of the TO 
and LO Raman lineshapes while using no adjustable parameter, 
and first-principles bond length calculations in the impurity 
limits and close to the bond percolation thresholds. In the latter 
case it is essential that we discriminate between isolated and 
connected bonds, not in the usual terms of next-nearest neighbors. 

The TO and LO modes in the Raman/IR spectra are re-assigned, 
with notable difference from previous attributions. In particular 
the GaP:In impurity mode, earlier identified within the GaP optical 
band, i.e. at $\sim$390~cm$^{-1}$, is re-assigned at a frequency 
significantly below the TO mode (368~cm$^{-1}$), i.e. at $\sim$350~cm$^{-1}$. 
Accordingly the two impurity modes stay out of the TO--LO bands 
of the host compounds, so that the In--P and Ga--P TO phonon branches 
do not overlap. This is enough to reconcile (Ga,In)P with the 
Elliott's criterion. 

In the TO symmetry the final picture consists of two well-separated 
Ga--P phonon branches just above two In--P branches so tight that 
they merge into a single overall In--P branch. On this basis the 
dominant TO mode in the Raman/IR spectra is re-assigned as a 
(In--P, Ga--P)-mixed mode resulting from the sum of the overall 
In--P mode and the near-by low-frequency Ga--P mode. Besides, 
the minor TO mode is re-assigned as the remaining high-frequency 
Ga--P mode. Regarding the LO symmetry we show that strong coupling 
occurs between the individual LO modes, via their long range 
longitudinal polarization field. The resulting LO signal is strongly 
distorted with respect to the individual uncoupled LO lines, 
which makes it not relevant to attribute any LO feature in the 
Raman/IR spectra to any specific bond vibration, at any alloy 
composition.

Basically our simple 1-bond$\rightarrow$2-phonon phenomenological model 
appears to provide consistent understanding of the long wave (TO, LO) 
phonon properties of random (Ga,In)P, which was still lacking.

Moreover, we propose a mechanism for spontaneous ordering in 
GaInP$_2$, based on our observation of a bi-modal phonon behavior 
for the short Ga--P bond in random (Ga,In)P. Our view is that 
spontaneous ordering tends to favor those local atomic arrangements 
around the Ga sites that eventually result in longer Ga-P bond 
length. This way the local strain energy in the crystal due to 
the bond length mismatch between the parent compounds is minimized. 
In particular this simple mechanism accounts for two puzzling 
behaviors in the Raman/IR spectra when ordering increases: the 
reinforcement of the dominant mode to the cost of the minor mode 
in the TO symmetry, and the reduction of the so-called valley-to-depth 
ratio in the LO symmetry. Besides, as an unprewied issue, we 
predict an intrinsic limit to spontaneous ordering in mixed crystals. 
In GaInP$_2$ this should be reached when the minor TO mode completely 
disappears from the Raman/IR spectra. Experimentally this corresponds 
to $\eta\sim$0.5. It is worth to mention that the attempts 
to understand why spontaneous ordering failed to generate 
$\eta$ values greater than 0.5 had attracted little attention so far, 
whereas higher degrees or spontaneous ordering were currently 
under expectation. 

Generally this work illustrates that detailed understanding of 
the phonon mode behavior in mixed crystals requires to take into 
account the disorder in the force constant on top of the mass 
disorder. As a matter of fact generalization of the Coherent 
Potential Approximation to include the disorder in the force 
constant as a full theory was recently achieved by Ghosh \textit{et 
al.}$^{36}$ and Alam and Mookerjee,$^{37}$ for example, but we are not 
aware of any application to optical properties. 

This work has been supported by the Indo-French Center for the 
Promotion of Advanced Research (IFCPAR project No 3204-1).

\section*{References}
\noindent
$^{~1}$ I.F. Chang and S.S. Mitra, Adv. Phys. \textbf{20}, 359 (1971). \\
$^{~2}$ I.F. Chang and S.S. Mitra, Phys. Rev. \textbf{172} 924. (1968). \\
$^{~3}$ R.J. Elliott, J.A. Krumhansl and P.L. Leath, Rev. Mod. Phys. 
       \textbf{46}, 465 (1974). \\
$^{~4}$ A. Mascarenhas, H.M. Cheong, M.J. Seong and F. Alsina, 
       in \textit{Spontaneous Ordering in Semiconductor alloys}, \\
       \hspace*{3mm}
       edited by A. Mascarenhas (Springer Physics, 2002), p. 391. \\
$^{~5}$ G. Lucovsky, M.H. Brodsky, M.F. Chen, R.J. Chicotka and A.T.  Ward, 
       Phys. Rev. B \textbf{4}, 1945 (1971). \\
$^{~6}$ R. Beserman, C. Hirlimann and M. Balkanski, Solid State Commun. 
       \textbf{20}, 485 (1976). \\
$^{~7}$ E. Jahne, W. Pilz, M. Giehler and L. Hildisch, 
       Phys. Stat.  Sol. (b) \textbf{91}, 155 (1979).\\
$^{~8}$ H.W. Verleur and A.S.Jr. Barker, Phys. Rev. \textbf{149}, 715 (1966).\\
$^{~9}$ H.W. Verleur and A.S.Jr. Barker, Phys. Rev. \textbf{155}, 750 (1967).\\ 
$^{10}$ B. Jusserand and S. Slempkes, Solid State Commun. \textbf{49}, 
       95 (1984). \\
$^{11}$ T. Kato, T. Matsumoto and T. Ishida, 
        Jpn. J. Appl. Phys. \textbf{27}, 983 (1988). \\
$^{12}$ V. Ozolin\v{s} and A. Zunger, Phys. Rev. B \textbf{57}, R9404 (1998).\\
$^{13}$ V. Ozolin\v{s} and A. Zunger, Phys. Rev. B \textbf{63}, 87202 (2001).\\
$^{14}$ P.H. Borcherds, G.F. Alfrey, D.H. Saunderson, and A.D.B.  Woods, 
        J. Phys. C: Sol. State Phys. \textbf{8}, 2022 (1975). \\
$^{15}$ P.H. Borcherds, K. Kunc, G.F. Alfrey and R.L. Hall, 
       J. Phys. C: Solid State Phys. \textbf{12}, 4699 (1979). \\
$^{16}$ O. Pag\`{e}s, M. Ajjoun, T. Tite, D. Bormann, E. Tourni\'{e} 
       and K.C. Rustagi, Phys. Rev. B \textbf{70}, 155319 (2004), \\
       \hspace*{3mm} and references therein. \\
$^{17}$ O. Pag\`{e}s, T. Tite, K. Kim, P.A. Graf, O. Maksimov and 
        M.C. Tamargo, J. Phys.: Condens. Matter \textbf{18}, 577 (2006). \\
$^{18}$ D. Stauffer, \textit{Introduction to Percolation Theory} (Taylor 
        and Francis, London, 1985). \\
$^{19}$ O. Pag\`{e}s, T. Tite, A. Chafi, D. Bormann, O. Maksimov and 
        M.C. Tamargo, J. Appl. Phys., to be published. \\
$^{20}$ L. Bellaiche, S.-H Wei and A. Zunger, Phys. Rev. B \textbf{54}, 
        17568 (1996). \\
$^{21}$ A.V. Postnikov, O. Pag\`{e}s and J. Hugel, Phys. Rev. B \textbf{71}, 
        115206 (2005). \\
$^{22}$ R.M. Martin, Phys. Rev. B \textbf{1}, 4005 (1970). \\
$^{23}$ P. Ordej\'{o}n, E. Artacho and J.M. Soler, Phys. Rev. B \textbf{53}, 
        R10441 (1996). \\
$^{24}$ J.M. Soler, E. Artacho, J.D. Gale, A. Garc\'{\i}a, J. Junquera, 
        P. Ordej\'{o}n and D. S\'{a}nchez-Portal, \\
	\hspace*{3mm} J. Phys.: Condens. Matter \textbf{14}, 2745 (2002). \\
$^{25}$ N. Troullier and J.L. Martins, Phys. Rev. B \textbf{43}, 1993 
       (1991). \\
$^{26}$ M. Cardona, P. Etchegoin, H.D. Fuchs and P. Molin\`{a}s-Mata, 
        J. Phys.: Condens. Matter \textbf{5}, A61 (1993). \\
$^{27}$ R. Trommer, H. M\"{u}ller and M. Cardona, Phys. Rev. B \textbf{21}, 
        4869 (1980). \\
$^{28}$ R. Carles, G. Landa and J.B. Renucci, Solid State Commun. \textbf{53}, 
        179 (1985). \\
$^{29}$ W.L. Faust, and C.H. Henry, 
        Phys. Rev. Lett. \textbf{17}, 1265 (1966). \\
$^{30}$ J.F. Young, and K. Wan, Phys. Rev. B \textbf{35}, 2544 (1987). \\
$^{31}$ J.C. Mikkelsen and J.B. Boyce, Phys. Rev. B \textbf{28}, 7130 (1983).\\
$^{32}$ F. Alsina, J.D. Webb, A. Mascarenhas, J.F. Geisz, J.M. Olson 
        and A. Duda, Phys. Rev. B \textbf{60}, 1484 (1999). \\
$^{33}$ N. Mestres, F. Alsina, J. Pascual, J.M. Bluet, J. Camassel, 
        C. Geng and F. Scholz, Phys. Rev. B \textbf{54}, 17754 (1996). \\
$^{34}$ A.M. Mintairov, J.L. Merz and A.S. Vlasov, Phys. Rev. B \textbf{67}, 
        2052 (2003). \\
$^{35}$ H.M. Cheong, F. Alsina, A. Mascarenhas, J.F. Geisz and J.M.  Olson, 
        Phys. Rev. B \textbf{56}, 1888 (1997). \\
$^{36}$ S. Ghosh P.L. Leath and M.H. Cohen, Phys. Rev. B \textbf{66}, 
        214206 (2002). \\
$^{37}$ A. Alam and A. Mookerjee, Phys. Rev. B \textbf{71}, 94210 (2005).

\end{document}